\documentclass[aps,noshowpacs,twocolumn,superscriptaddress]{revtex4-1}  
\usepackage{graphicx}  
\usepackage{dcolumn}   
\usepackage{bm}        
\usepackage{ucs}
\usepackage{amssymb}   
\usepackage{amsmath}
\usepackage{amsfonts}
\usepackage{gensymb}
\usepackage{bibunits}
\usepackage{siunitx}
\usepackage{braket} 
\usepackage{caption,setspace}

\defaultbibliography{flatten_hybrid_pmixing}
\defaultbibliographystyle{ieeetr}

\begin{document}



\title{Electrically tunable organic-inorganic hybrid polaritons with monolayer WS$_2$}

\author{L.~C.~Flatten} \email{lucas.flatten@materials.ox.ac.uk} \affiliation{Department of Materials, University of Oxford, Parks Road, Oxford OX1 3PH, United Kingdom} 
\author{D.~M.~Coles} \affiliation{Department of Materials, University of Oxford, Parks Road, Oxford OX1 3PH, United Kingdom} \affiliation{Clarendon Laboratory, Department of Physics, University of Oxford, OX1 3PU, United Kingdom}
\author{Z.~He} \affiliation{Department of Materials, University of Oxford, Parks Road, Oxford OX1 3PH, United Kingdom}
\author{D.~G.~Lidzey} \affiliation{Department of Physics \& Astronomy, University of Sheffield, Sheffield S3 7RH, United Kingdom}
\author{R.~A.~Taylor} \affiliation{Clarendon Laboratory, Department of Physics, University of Oxford, OX1 3PU, United Kingdom}
\author{J.~H.~Warner} \affiliation{Department of Materials, University of Oxford, Parks Road, Oxford OX1 3PH, United Kingdom}
\author{J.~M.~Smith} \email{jason.smith@materials.ox.ac.uk} \affiliation{Department of Materials, University of Oxford, Parks Road, Oxford OX1 3PH, United Kingdom}
\date{\today}

\begin{bibunit}

\begin{abstract}
Exciton-polaritons are quasiparticles consisting of a linear superposition of photonic and excitonic states, offering potential for nonlinear optical devices. The excitonic component of the polariton provides a finite Coulomb scattering cross section, such that the different types of exciton found in organic materials (Frenkel) and inorganic materials (Wannier-Mott) produce polaritons with different interparticle interaction strength. A hybrid polariton state with distinct excitons provides a potential technological route towards in-situ control of nonlinear behaviour. Here we demonstrate a device in which hybrid polaritons are displayed at ambient temperatures, the excitonic component of which is part Frenkel and part Wannier-Mott, and in which the dominant exciton type can be switched with an applied voltage. The device consists of an open microcavity containing both organic dye and a monolayer of the transition metal dichalcogenide WS$_2$. Our findings offer a perspective for electrically controlled nonlinear polariton devices at room temperature.
\end{abstract}

\maketitle
\section*{\label{Introduction} Introduction}
Exciton-polaritons result from strongly coupling an optical cavity mode to electronic transitions with reversible energy exchange between the two \cite{weisbuch_observation_of_the,lidzey_strong_exciton_photon}. Polaritons inherit properties such as the delocalised photonic wavefunction and finite excitonic mass from their constituents. Nonlinearities in an ensemble of polaritons arise as a result of the Coulomb interaction via the excitonic fraction of the state \cite{dang_stimulation_of_polariton,senellart_nonlinear_emission_of}. Since the first demonstration of exciton-polaritons \cite{weisbuch_observation_of_the}, strongly correlated effects such as Bose-Einstein condensation in a semiconductor quantum well microcavity at cryogenic temperatures \cite{kasprzak_bose_einstein_condensation} and more recently with a polymer at room temperature \cite{plumhof_room-temperature_2014} have been shown. Superfluidity \cite{amo_superfluidity_of_polaritons}, polariton lasing \cite{schneider_electrically_2013} and multistability \cite{rodriguez_interaction-induced_2016} have furthered the range of nonlinear phenomena these ultra-light bosonic quasiparticles display. The exciton part of polaritons plays a vital role as it conveys the interparticle interaction potential. Generally two different types of excitons are distinguished. The Frenkel exciton is characterized by its strong binding energy (order of \SI{1}{\electronvolt}) and large oscillator strength \cite{lidzey_room_temperature_polariton}. However, a small Bohr radius ($\sim$~\SI{1}{\nano\meter}) and relatively low mobility ($\sim 10^{-2}$~\SI{}{\centi\meter}$^{2}$\SI{}{\volt}$^{-1}$\SI{}{\second}$^{-1}$) make for weak exciton-exciton interaction cross sections. Wannier-Mott excitons generally have lower binding energies (a few \SI{}{\milli\electronvolt}) and smaller oscillator strengths but larger scattering cross sections. For the recently emerged class of two-dimensional transition metal dichalcogenides (TMD) the exciton classification is not trivial since the exciton binding energy can be as high as \SI{700}{\milli\electronvolt} \cite{zhu_exciton_2015}. However, due to their crystalline structure the carrier mobility is large ($\sim 40$~\SI{}{\centi\meter}$^{2}$\SI{}{\volt}$^{-1}$\SI{}{\second}$^{-1}$) \cite{schmidt_electronic_2015}, the Coulomb interaction in the confined plane is strong and the wavefunction is of Wannier-Mott type (i.e. with an extension over a large number of unit cells)\cite{liu_electronic_2015}. TMD monolayers are direct bandgap materials that display intriguing optical properties potentially useful for future applications and fundamental research, such as high quantum yields and strong absorption in the visible, enhanced excitonic Coulomb interactions and valley degrees of freedom \cite{ross_electrical_2013,schmidt_electronic_2015,liu_electronic_2015,allain_electrical_2015}. Through a proper choice of substrate and the reduction of impurities the mobilities in such monolayers can be enhanced dramatically and electrical charge injection is possible \cite{klots_probing_2014, allain_electrical_2015}. 

\begin{figure*}
\includegraphics[width=0.8\textwidth]{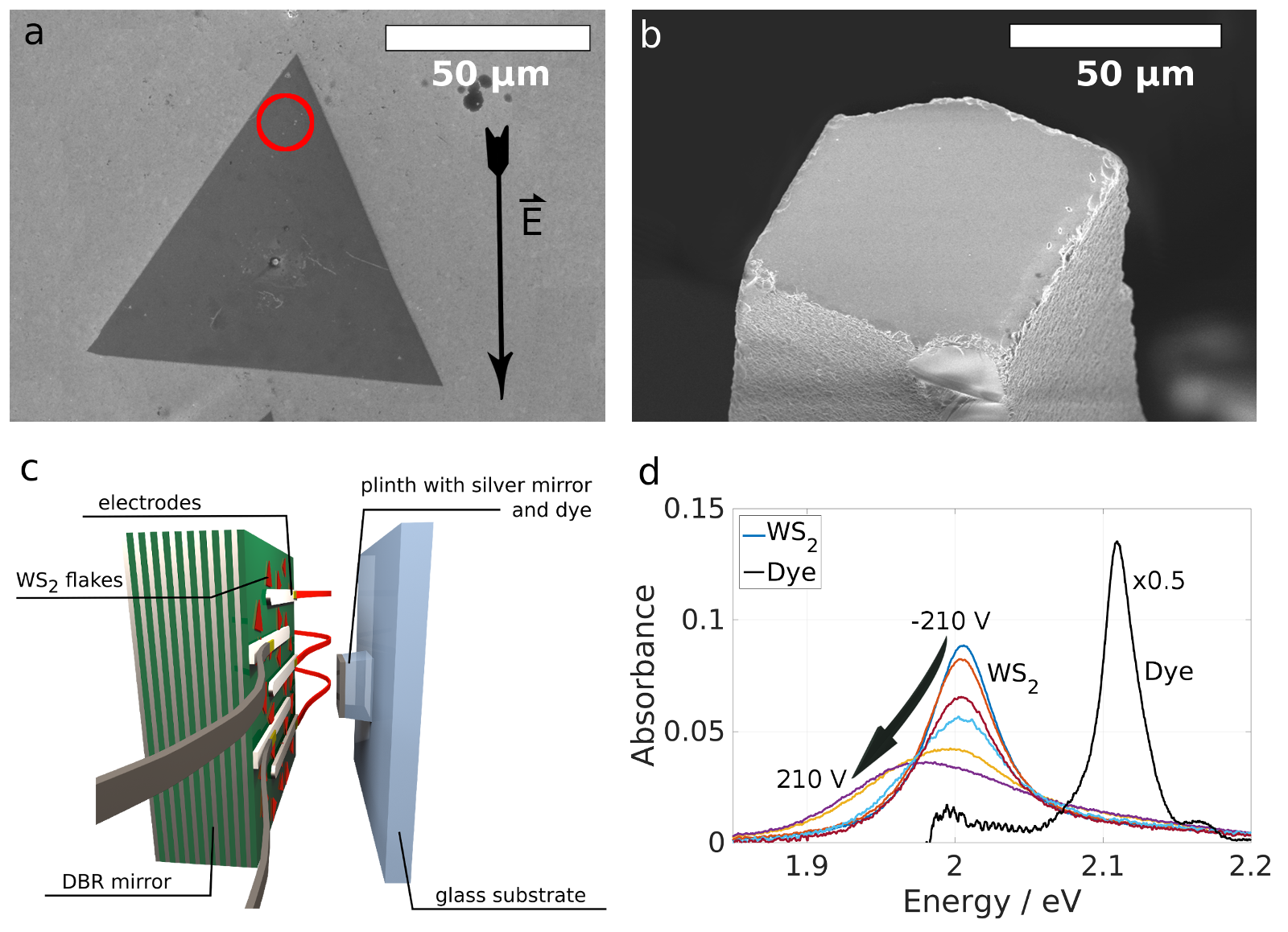}
\caption[flushleft]{\textbf{Two-dimensional WS$_2$ between electrodes in an optical microcavity.} a) SEM-micrograph of a monolayer WS$_2$ flake deposited on SiO$_2$ terminated DBR, forming one side of the optical microcavity. The arrow denotes the direction of the applied electric field and the red circle the area from which the data was obtained. b) SEM image of the opposing cavity side, a silver mirror on a silica plinth. c) Sketch of the two mirrors with silver electrodes on the surface of the DBR giving electrical tunability within the cavity. d) Absorbance of atomically-thin WS$_2$ for various applied voltages obtained from the position marked with a red circle in a (colour) and the organic dye TDBC (black, scaled by factor 0.5).
\label{fig1} }
\end{figure*}

Engineering polariton states for nonlinear effects in compact optoelectronic devices remains a major challenge. A possible pathway towards controlling the physical properties of polaritons within a device is the use of hybrid polariton states \cite{holmes_strong_coupling_and,wenus_hybrid_organic_inorganic}. Such states combine excitonic properties from different materials, thus creating polaritons with properties beyond those that can be created by any single material. The superposition weight of their components can be modified in-situ by changing the resonance condition between the different excitons and the cavity mode. 

In this article we demonstrate the formation of hybrid organic-inorganic polaritons created through the simultaneous coupling of the J-aggregate dye TDBC and a tungsten-disulphide (WS$_2$) monolayer to a confined optical microcavity mode \cite{rong_controlling_2014,flatten_room-temperature_2016}. The cavity consists of a distributed Bragg reflector (DBR) with WS$_2$ flakes on the low refractive index terminated side and a small silver mirror covered with a thin layer of the organic dye (Fig. \ref{fig1}a-c). When the cavity mode energy is tuned between the two exciton energies, it couples simultaneously to both exciton types thereby creating hybrid organic-inorganic polariton states. The WS$_2$ layer is placed between two electrodes such that an applied voltage perturbs the energy of the Wannier-Mott excitons and thereby modifies the composition of the hybrid polariton state. The change in the spectral position alters the relative mixing of Frenkel and Wannier-Mott excitons within polariton states, thus allowing for in situ control of polariton properties such as mobility and scattering cross section.

\section*{\label{results} Results}
\subsection*{Hybrid polariton states}
\begin{figure*}
\includegraphics[width=1.0\textwidth]{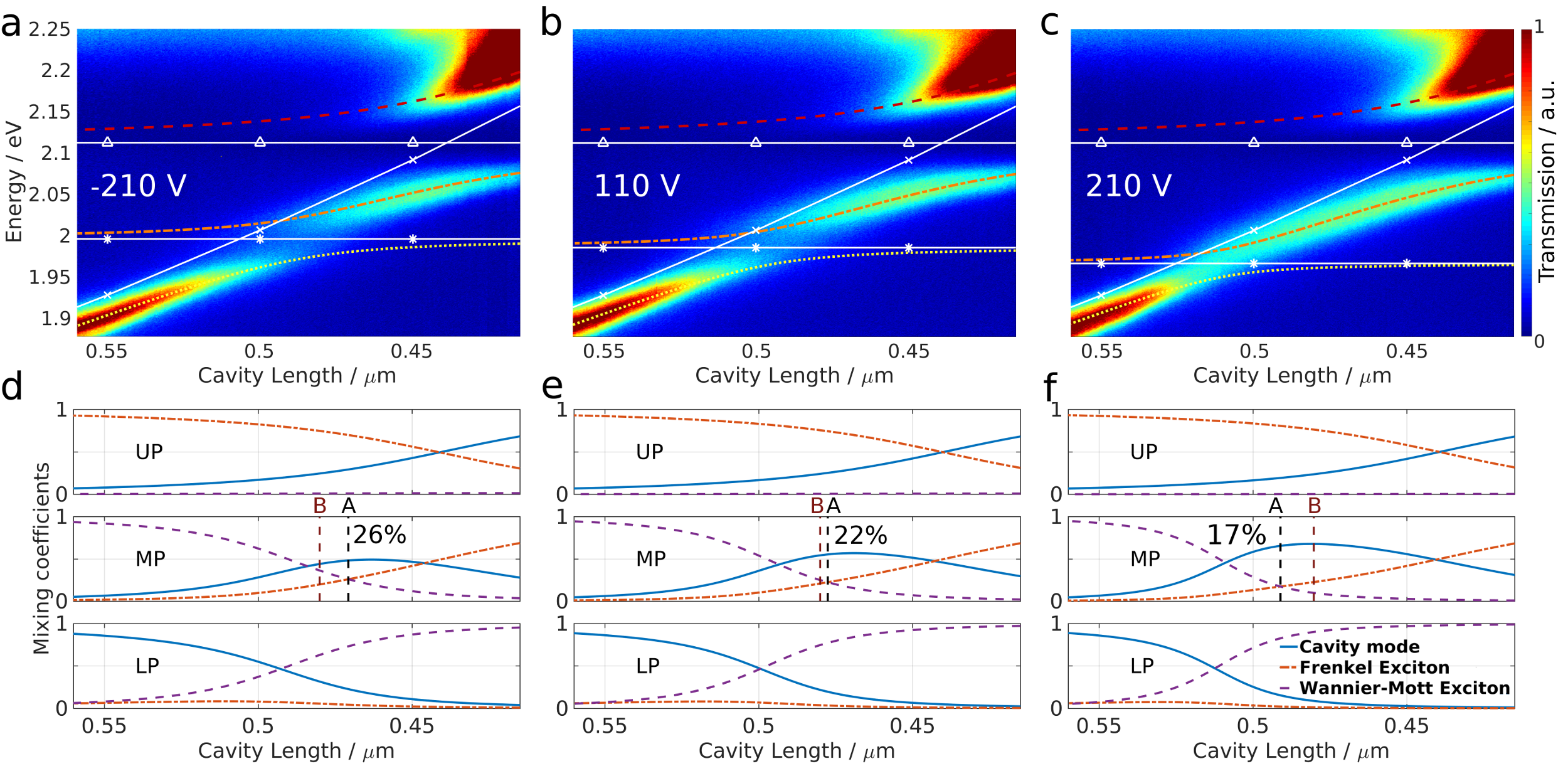}
\caption[flushleft]{\textbf{Electrically controlled hybridisation of Frenkel- and Wannier-Mott-excitons in a polariton state.} a-c) Successive transmission spectra of hybrid WS$_2$-TDBC microcavity for decreasing cavity length from left to right and different applied voltages of \SI{-210}{\volt}, \SI{110}{\volt} and \SI{210}{\volt} for a, b and c respectively. The white, continuous lines correspond to the uncoupled energies of Frenkel-exciton (TDBC, triangles), Wannier-Mott exciton (WS$_2$, stars) and cavity mode (crosses). The dashed lines in colour show the dispersion for the coupled system consisting of the three polariton branches, lower polariton (LP, yellow, - -), middle polariton (MP, orange, $-$ -) and upper polariton (UP, red, $-$ $-$), where the terms in brackets denote the name of the state, the line colour and the line style respectively. d-f) Photonic (cavity mode, blue, continuous), Frenkel-excitonic (TDBC, red,$-$ -) and Wannier-Mott-excitonic (WS$_2$, purple, - -) contribution to the three polariton branches LP, MP and UP for the dispersions plotted above respectively. Two points A (black dashed line) and B (red dashed line) mark cavity lengths at which: A) Frenkel- and Wannier-Mott-exciton contribution to the middle polariton branch is equal, B) the cavity length is $L=$~\SI{0.48}{\micro\meter} and the dominant exciton contribution can be swapped electrically. The numerical value displayed in the MP panel as a percentage gives the value of Frenkel- and Wannier-Mott-exciton contribution $\beta^2$ and $\gamma^2$ at point A. \label{fig2} }
\end{figure*}

Both the organic dye and the inorganic WS$_2$ monolayer absorb strongly in the visible, with narrow, excitonic absorption peaks at $E_{\rm{F}}=$~\SI{2.11}{\electronvolt} and $E_{\rm{WM}}=$~\SI{2.01}{\electronvolt} respectively. Placed in an electric field, the WS$_2$ absorption peak position is additionally tunable within approximately \SI{20}{\milli\electronvolt} to the red for electric field strengths of $E =2.33\times 10^4$~\SI{}{\volt}\SI{}{\centi\meter}$^{-1}$ (see Fig.~\ref{fig1}d). This effect is caused by changes in the local electron density which causes a shift in spectral weight of neutral (X$^0$) and charged excitons (X$^-$) and has been discussed elsewhere \cite{mak_tightly_2013,ross_electrical_2013,plechinger_identification_2015,chernikov_electrical_2015}. Here we use the effect to change the WS$_2$ absorption peak position, amplitude and width within the cavity to alter a polariton state. This polariton state is of hybrid nature, combining two different species of excitons, Wannier-Mott (WM) excitons  formed in the inorganic component WS$_2$ and Frenkel (F) excitons in the organic J-aggregated dye TDBC. The system of one cavity mode simultaneously coupled to two excitonic transitions is described by the Hamiltonian 
\begin{equation}
\begin{split}
H = \  & E_c b^\dagger b
+ E_{\rm{F}} x_{\rm{F}}^\dagger x_{\rm{F}}
+ E_{\rm{WM}} x_{\rm{WM}}^\dagger x_{\rm{WM}} +\\ 
& \  V_{\rm{F}} (b^\dagger x_{\rm{F}} + b x_{\rm{F}}^\dagger  )
+ V_{\rm{WM}} (b^\dagger x_{\rm{WM}} + b x_{\rm{WM}}^\dagger  )
\end{split}
\label{eq2}
\end{equation}
where $V_{\rm{F}}$ and $V_{\rm{WM}}$ are the interaction potentials between the cavity mode and F- and WM excitons. $b, x_{\rm{F}}$ and $x_{\rm{WM}}$ are the photon, F- and WM-exciton annihilation operators respectively (here for one $k$ vector only). In the stationary case the system can be reduced to:
\begin{equation}
H \ket{\Psi} 
=
\begin{pmatrix} 
E_{\rm{c}} & V_{\rm{F}} & V_{\rm{WM}} \\
V_{\rm{F}} & E_{\rm{F}} & 0 \\
V_{\rm{WM}} & 0 & E_{\rm{WM}} 
\end{pmatrix}
\begin{pmatrix} 
\alpha \\
\beta \\
\gamma
\end{pmatrix} = E \ket{\Psi} 
\label{eq1}
\end{equation}
Here the state $\ket{\Psi}$ is defined by the three coefficients $\alpha$, $\beta$ and $\gamma$, which quantify the contribution of photon, F- and WM-exciton respectively. Each of the three eigenstates forms a polariton branch, whose photonic part can be readily observed spectroscopically. Fig.~\ref{fig2}a-c show the dispersion of these branches for different applied voltages of -210 V, 110 V and 210 V respectively, as measured by taking successive transmission spectra for varying cavity lengths. As the cavity length is decreased from $L=$~\SI{0.56}{\micro\meter} to $L=$~\SI{0.41}{\micro\meter} the cavity mode energy increases according to $E_{\rm{c}} = \frac{q h c}{2 L}$, where $q=4$ is the longitudinal mode index (white continuous line, $\times$). As it traverses the exciton energies at $E_{\rm{WM}}$ (white continuous line,  *) and $E_{\rm{F}}$ (white continuous line,  $\Delta$), an anti-crossing is visible which is indicative of the strongly coupled nature of the system. In this way, coupled eigenstates to Eq. \ref{eq1} are formed which we call the lower, middle and upper polariton branch (LP, MP and UP). The constituent uncoupled components for these branches quantified by $\alpha^2$, $\beta^2$ and $\gamma^2$ are shown in Fig.~\ref{fig2} d-f. The nature of the LP (UP) branch transits from photonic (F-excitonic) to WM-excitonic (photonic) as the cavity length is decreased. The more interesting MP branch changes from WM-excitonic to F-excitonic nature for decreasing mirror separation. For cavity lengths in the region \SI{0.525}{\micro\meter}~$> L >$~\SI{0.425}{\micro\meter}, its photonic fraction increases and together with the two excitons, a hybrid polariton state is formed. We shall define two points of interest along the dispersion curve: Point A corresponds to the cavity length at which the middle polariton branch has equal weights of F- and WM-exciton, and is labelled in Fig.~\ref{fig2}d-f (middle panels). Point B stands for a fixed cavity length of $L=$~\SI{0.480}{\micro\meter}, which we will show corresponds to a cavity length where we can electrically switch the MP state from having a dominant Wannier-Mott exciton component to a dominant Frenkel exciton component.

\subsection*{Electrical control over polariton composition}
By applying an external electric field, $E_{\rm{WM}}$ and the corresponding absorption amplitude and linewidth are modified leading to a change in the polariton mixing coefficients. As the applied voltage is increased from \SI{-210}{\volt} to \SI{110}{\volt} (\SI{210}{\volt}), the Rabi splitting about the Wannier-Mott exciton decreases from $2 V_{\rm{WM}} =$ \SI{57}{\milli\electronvolt} to \SI{46}{\milli\electronvolt} (\SI{35}{\milli\electronvolt}) and $E_{\rm{WM}}$ shifts from \SI{1.997}{\electronvolt} to \SI{1.987}{\electronvolt} (\SI{1.968}{\electronvolt}). The absorption linewidth of the WM-excitonic transition increases such that the splitting is not resolvable for applied voltages above \SI{150}{\volt} (see Fig. \ref{fig2}c). The Rabi splitting about the Frenkel exciton energy is $2 V_{\rm{F}} =$ \SI{114}{\milli\electronvolt} for all applied voltages. For the same change in voltage the excitonic weight at point A changes from $\beta^2 = \gamma^2 = 26\%$ to $22\%$ ($17\%$) while the photonic fraction increases from $\alpha^2 = 48\%$ to $56\%$ ($66\%$). At the same time the cavity length corresponding to point A shifts from $L=$~\SI{0.471}{\micro\meter} to \SI{0.478}{\micro\meter} (\SI{0.492}{\micro\meter}).
At point B the composition of the polariton swaps from $\beta^2 = 20\%$, $\gamma^2 = 36\%$ to $\beta^2 = 21\%$, $\gamma^2 = 25\%$ ($\beta^2 = 23\%$, $\gamma^2 = 10\%$) while $\alpha^2$ increases from $44\%$ to $54\%$ ($67\%$) as the applied voltage is changed from \SI{-210}{\volt} to \SI{110}{\volt} (\SI{210}{\volt}), therefore the dominant exciton component within the MP branch switches from Wannier-Mott to Frenkel with increasing voltage. 
\begin{figure*}
\includegraphics[width=0.8\textwidth]{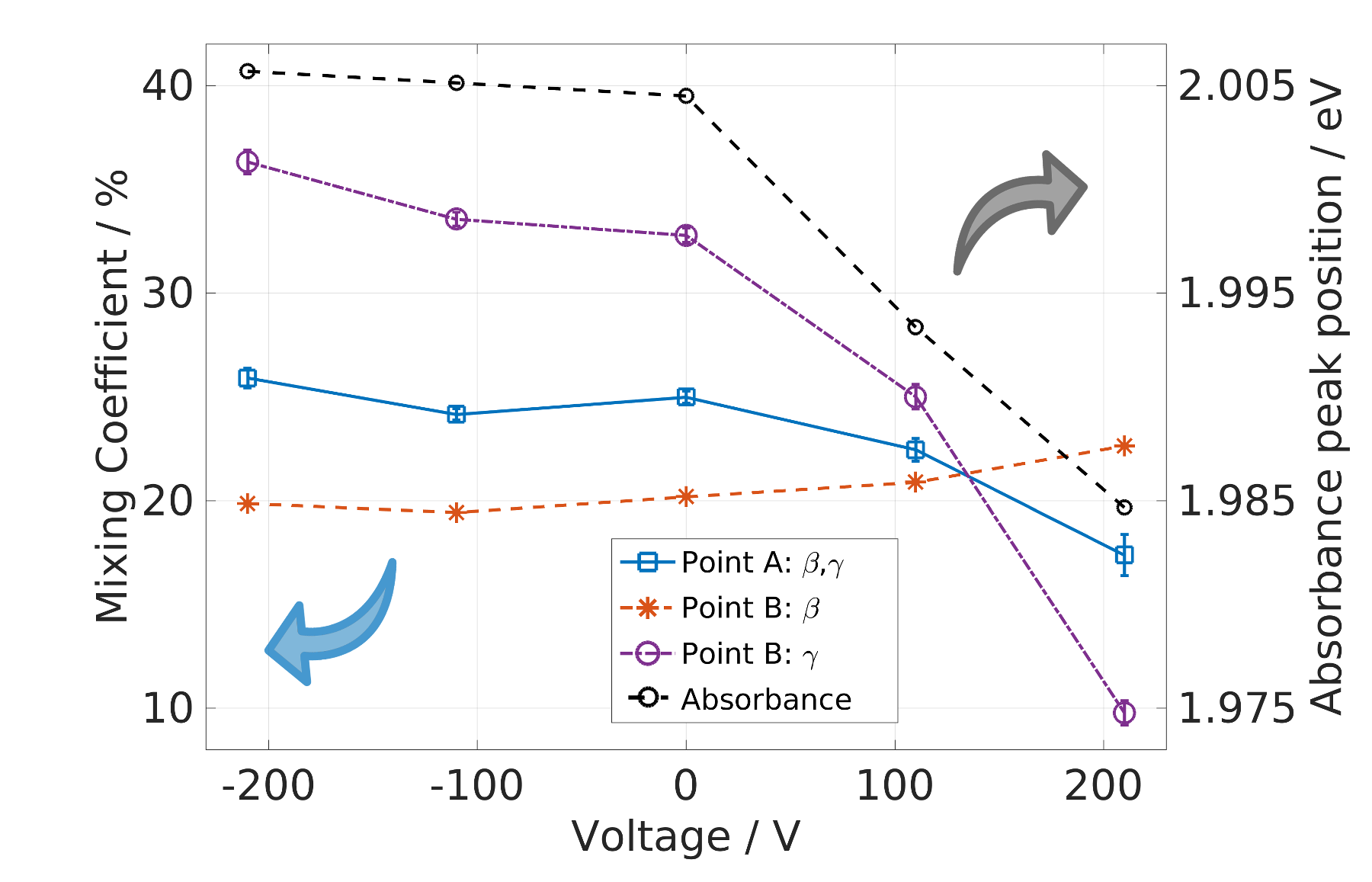}
\caption[flushleft]{\textbf{Electrical control over polariton composition.} Frenkel- and Wannier-Mott-exciton fraction $\beta$ and $\gamma$ in hybrid polariton state for different applied voltages (colour, symbols, left ordinate, blue arrow) and interpolated absorbance peak position of WS$_2$ outside the cavity (black, dashed, right ordinate, grey arrow). The exciton fractions are shown for two points, where point A corresponds to the cavity length at which $\beta = \gamma$ (maximal mixing) and point B to a fixed cavity length of $L=$~\SI{0.48}{\micro\meter}. The errorbars were obtained by fitting the polariton dispersion (Eq. \ref{eq1}) to the transmission data and are of similar size as the symbols. The absorbance peak positions were acquired by fitting a Gaussian lineshape to the absorbance shown in Fig.~\ref{fig1}. More information about the error estimation is given in Supplementary Fig. 6 and Supplementary Note 4. \label{fig3} }
\end{figure*}
Fig.~\ref{fig3} shows the changes in polariton composition as a function of the applied voltage for more intermediate voltages. The black dashed line is obtained by fitting the absorption peaks presented in Fig.~\ref{fig1}d with a Gaussian lineshape and interpolating the central energy (right ordinate). The datapoints quantifying the polariton composition are the result of fits of the polariton dispersion given by Eq. \ref{eq1} to the transmission data shown in Fig. \ref{fig2}a-c (see Supplementary Fig. 6). More details on the fitting procedure and the origin of the errorbars is given in Supplementary Note 4. It is evident that the change in the hybrid polariton composition stems from the altered WM-exciton state, which can be controlled electrically in the above stated manner. We present another dataset acquired from a different WS$_2$ flake in the Supplementary Note 5 and Supplementary Figure 7 and 8.

\section*{\label{conclusion} Discussion}
The change in absorption peak position and lineshape in response to the change in applied voltage originates from the local change in electron density. This effect has been described previously for the PL lineshape \cite{ross_electrical_2013,
plechinger_identification_2015,
zhu_exciton_2015} and more recently for the absorption spectrum of TMDs \cite{chernikov_electrical_2015,sidler_fermi_2016}. It is attributed to a combined result of Coulomb scattering, Pauli blocking and Coulomb screening which causes a transfer of oscillator strength from neutral (X$^0$) to charged (X$^-$) exciton together with a shift in energy of both states (see \cite{chernikov_electrical_2015,sidler_fermi_2016} for a more detailed description). In our system with a laterally applied field without direct carrier injection we make use of the abundance of electrons commonly found in WS$_2$ and MoSe$_2$ flakes \cite{ross_electrical_2013,
plechinger_identification_2015}, whose distribution across the flake can be altered with the electric field (see Supplementary Figure 3 and Supplementary Note 3). Due to the geometry of our system, the steady state of this distribution is reached on a millisecond timescale. We attribute this slow speed to the occurrence of scattering defects in the transferred monolayer, caused by locally induced strain and impurities reducing the mobility. Improvements in sample preparation such as the embedding of single TMDC layers within quasi non-interacting hBN heterostructures together with a back-gated electric field geometry would allow for fast switching times, limited with the current technology by the high resistance and therefore the high RC constant of the measurement  circuit \cite{klots_probing_2014}. Advances in contacting two-dimensional semiconductor layers up to the ohmic contact would remedy this situation \cite{allain_electrical_2015}.

While the transition from X$^0$ to X$^-$ with increased electron density is well understood \cite{mak_tightly_2013,chernikov_electrical_2015}, the impact such change has on exciton-polariton states is non-trivial and the topic of current research \cite{baeten_many-body_2015,sidler_fermi_2016}. Due to the fermionic nature of a trion state, the trion-trion interaction is stronger than the purely exciton mediated non-linearity \cite{portella-oberli_nonlinear_2003}, which would make trion-polariton systems attractive for observing strongly correlated phenomena obeying Fermi-Dirac statistics.

In addition to the electrical tuning of the polariton state, the open cavity allows to bring higher longitudinal modes with decreasing exciton-photon coupling into resonance \cite{flatten_room-temperature_2016}, which results in different mixing of the polariton components.

We have fabricated an open microcavity system containing a monolayer of transition metal dichalcogenide and layer of J-aggregate dye that are simultaneously strongly coupled to a cavity mode at room temperature. The resulting hybrid polariton states have a mixed Frenkel/Wannier-Mott exciton nature. Application of a transverse electric field across the monolayer results in a shift of the absorption peak energy and allows controllable tuning of the exciton mixing within the hybrid polariton states. Polariton-polariton interactions give rise to non-linear effects, rendering polaritonic systems attractive to observe a multitude of fascinating phenomena such as inversionless lasing, superfluidity and topologically non-trivial states \cite{fraser_physics_2016}. These interactions are much weaker for localised Frenkel excitons than for Wannier-Mott excitons typical in a crystalline lattice. Our results show how the hybridisation between such distinct excitons could be controlled electrically at room temperature. These findings could open pathways to novel photonic devices with engineered optical properties.

\section*{\label{setup} Methods}
\subsection*{Sample preparation}
The open microcavity consists of two opposing flat mirrors, a large dielectric distributed Bragg reflector (DBR) with 10 pairs of SiO$_2$, TiO$_2$ with central wavelength of $\lambda = $ \SI{640}{\nano\meter} and a smaller silver mirror (Fig.~\ref{fig1}b). To enable electrical control within the cavity, silver electrodes with a width and a spacing of \SI{90}{\micro\meter} are thermally evaporated on top of the DBR (see Supplementary Fig. 1). The thickness of these electrodes is similar to the thickness of the silver layer on the small opposing mirror,  approximately \SI{50}{\nano\meter}. The WS$_2$ flakes are grown as described in \cite{rong_controlling_2014} and transferred onto the dielectric mirror stack, which has a low refractive-index terminated configuration to provide an anti-node of the electric field at the mirror surface and thus optimal coupling to the monolayer (see Supplementary Fig. 2). The distribution of WS$_2$ flakes relative to the electrodes is random, with 90$\%$ of the flakes overlapping partially with the silver electrodes. The results presented here were obtained from a flake which was not in physical contact with the electrodes, thus ensuring purely electrostatic tuning (see Supplementary Figure 3). 
The other component of the hybrid system is the organic J-aggregated dye TDBC. After dissolving in a gelatine-water solution, the dye was spin-coated onto the small silver mirror to give a polymer-dye layer of about \SI{300}{\nano\meter} thickness. The absorbance of the TDBC film can be tuned by varying the concentration of the dye (see Supplementary Fig. 4 and 5). For a more detailed exposition of the sample preparation we refer the reader to the Supplementary Note 1a.

\subsection*{Measurements}
The small silver mirror with the dye was mounted on a three-dimensional piezo actuated stage, which makes positioning of the silver mirror relative to the WS$_2$ flake possible and allows electrical control of the cavity length.  By moving the silver plinth over a region of the DBR mirror which holds monolayer WS$_2$ and reducing the distance between the two mirrors below $L~\approx$~\SI{5}{\micro\meter}, stable cavity modes interacting strongly with the WS$_2$ excitons appear. The mode structure of the system can be observed by measuring the transmitted light of a spectrally broad lightsource. For this analysis the light was focused onto an Andor combined spectrometer/CCD. For more details on the measurements we refer the reader to the Supplementary Note 1b. Additionally we discuss the optical properties of the individual components of the hybrid polariton system in Supplementary Note 2.

\subsection*{Data availability}
The data that supports the findings of this study are available at the Oxford University Research Archive (https://ora.ox.ac.uk/).

\section*{Acknowledgements}
We thank Radka Chakalova at the Begbroke Science Park for helping with the thermal evaporation and dicing of the mirrors. L.F. acknowledges funding from the Leverhulme Trust. J.M.S. and D.M.C. acknowledge funding from the Oxford Martin School and EPSRC grant EP/K032518/1. R.A.T. also acknowledges funding from the Oxford Martin School. D.G.L. and D.M.C. acknowledge EPSRC grant EP/M025330/1.

\section*{Author contributions}
Z.H. and J.W. grew the WS$_2$ and deposited it on the mirror. D.C. prepared the organic dye, measured the Rabi splitting as a function of the concentration and coated the small mirror with it. All other sample preparation steps and measurements were carried out by L.F. under the supervision of R.T. and J.S.. All authors contributed to the preparation of the manuscript.

\section*{Additional Information}
This paper is accompanied by Supplementary Information.

Competing financial interests: The authors declare no competing financial interests.
\putbib

\end{bibunit}

\onecolumngrid
\newpage

\clearpage
\pagebreak
\widetext
\appendix
\begin{center}
\textbf{\large Supplemental Material: Electrically tunable organic-inorganic hybrid polaritons with monolayer WS$_2$}
\end{center}
\setcounter{equation}{0}
\setcounter{figure}{0}
\setcounter{table}{0}
\setcounter{page}{1}
\makeatletter
\renewcommand{\theequation}{S\arabic{equation}}
\renewcommand{\thefigure}{S\arabic{figure}}
\renewcommand{\bibnumfmt}[1]{[S#1]}
\renewcommand{\citenumfont}[1]{S#1}
\begin{bibunit}
\begin{figure}[h!]
\centering
\includegraphics[width=0.95\textwidth]{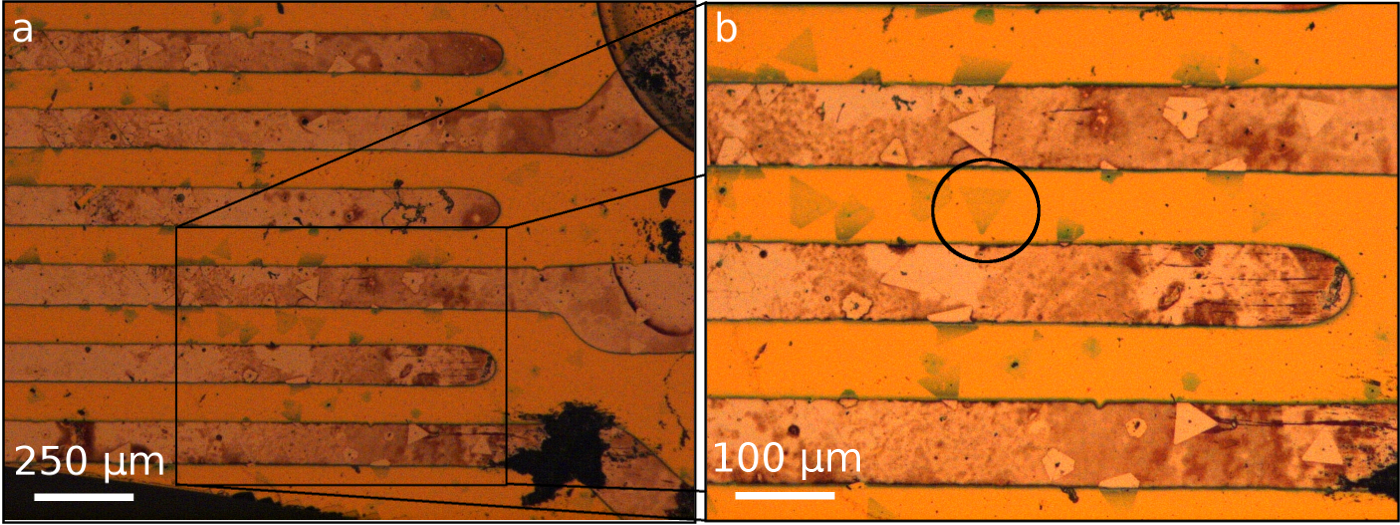}
\caption[flushleft]{\textbf{Optical microscope images of WS$_2$ flakes on DBR with electrodes.} a) Large field of view (x20) showing the individual silver electrodes, which are connected to either side. b) Zoom (x50) into region marked in a, showing randomly aligned WS$_2$ flakes on top of the silver electrodes. The circle marks one of the flakes used for this study. \label{figS7} }
\end{figure}

\begin{figure}[h!]
\centering
\includegraphics[width=1.0\textwidth]{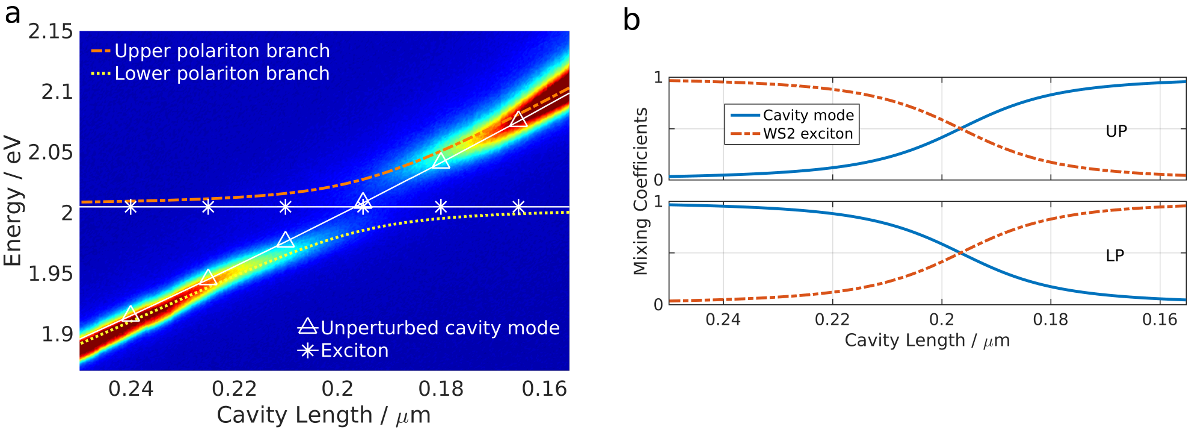}
\caption[flushleft]{\textbf{Polaritons with the inorganic component WS$_2$ only.} a) Successive transmission spectra as the cavity length is decreased from left to right. The unperturbed cavity mode (white continuous, $\Delta$) has the longitudinal mode index $q=3$. As it traverses the WS$_2$ exciton energy, the strong coupling is evidenced by the formation of lower (yellow, dashed) and upper (orange, dashed) polariton branches with a Rabi splitting of \SI{70}{\milli\electronvolt}. b) Photonic (blue, continuous) and excitonic (red, dashed) fractions of lower and upper polariton branch as the cavity length is swept. 
 \label{figS2}} 
\end{figure}

\begin{figure}[h!]
\centering
\includegraphics[width=0.65\textwidth]{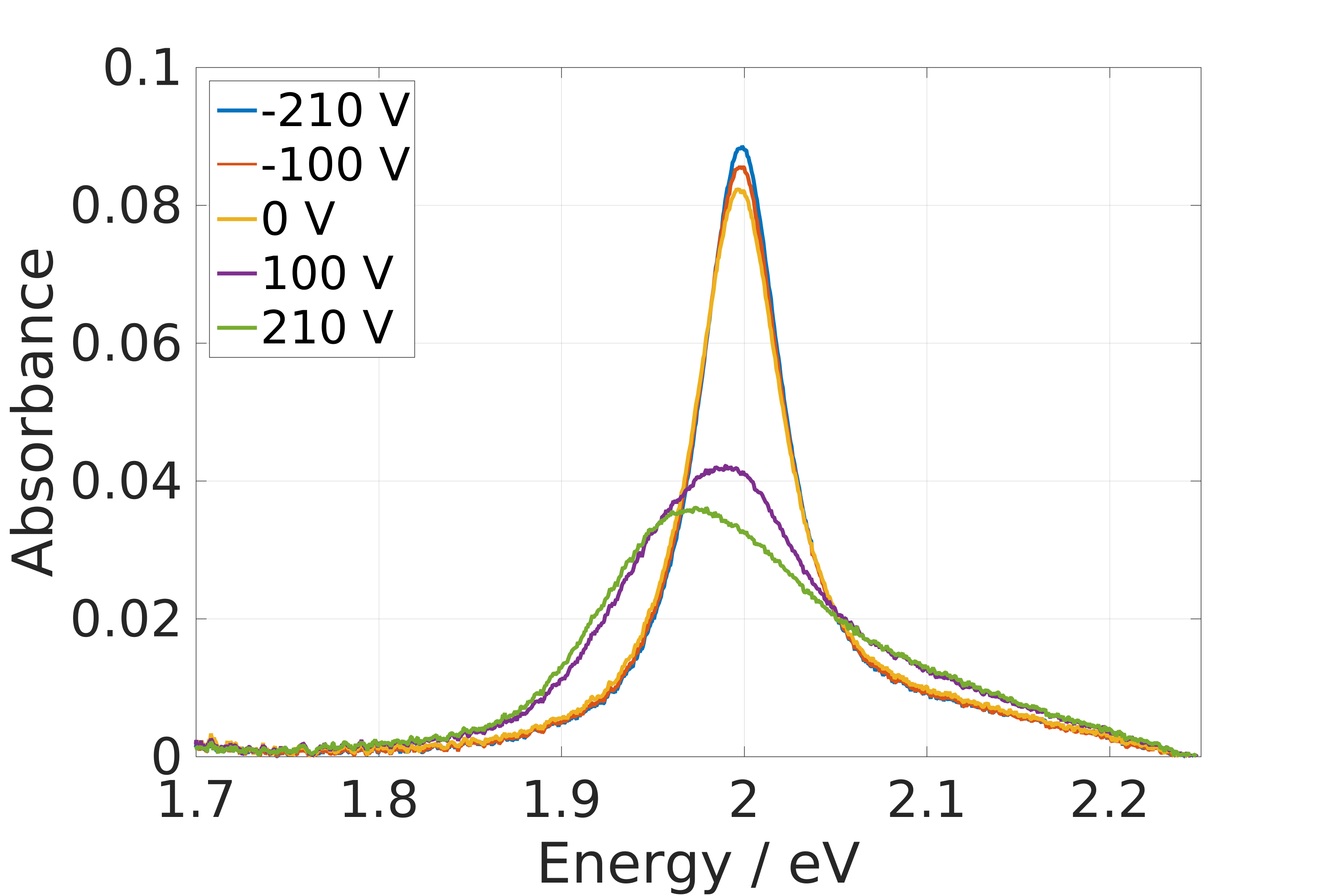}
\caption[flushleft]{\textbf{Electrically changed absorption profile.} WS$_2$ absorption profile as a function of the applied voltage. For large negative voltages the neutral exciton X$^0$ state dominates the profile (depletion of free electrons), while for the opposite bias the spectral weight shifts towards the charged exciton X$^-$ (abundance of free electrons).
 \label{figS5}} 
\end{figure}

\begin{figure}[h!]
\centering
\includegraphics[width=0.5\textwidth]{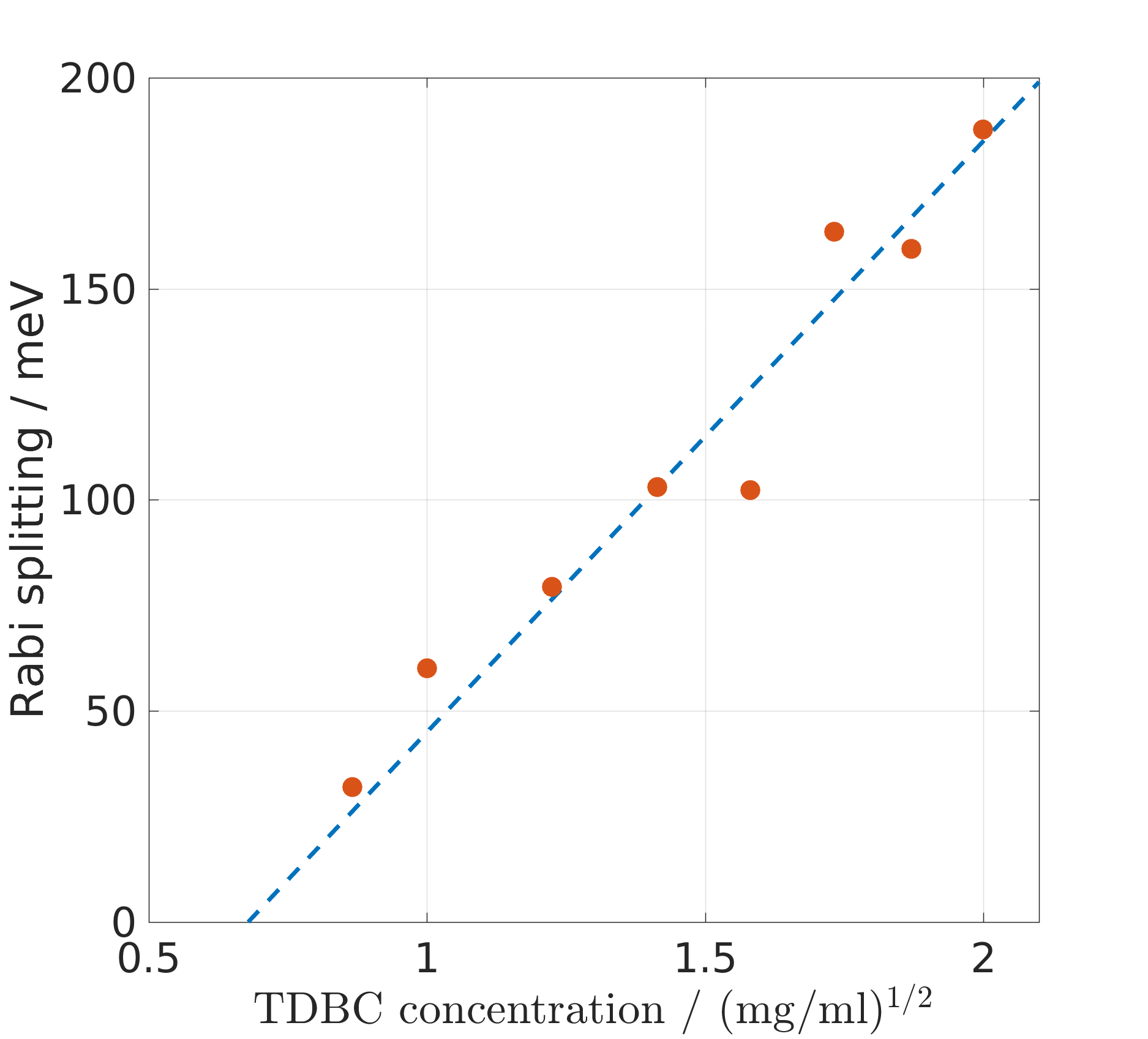}
\caption[flushleft]{\textbf{Polaritons with the organic component TDBC only.} Rabi splitting as a function of the square root of the TDBC concentration for a fixed cavity length, revealing a linear dependence when plotted against the square root of the dye concentration.
 \label{figS6}} 
\end{figure}

\begin{figure}[h!]
\centering
\includegraphics[width=1.0\textwidth]{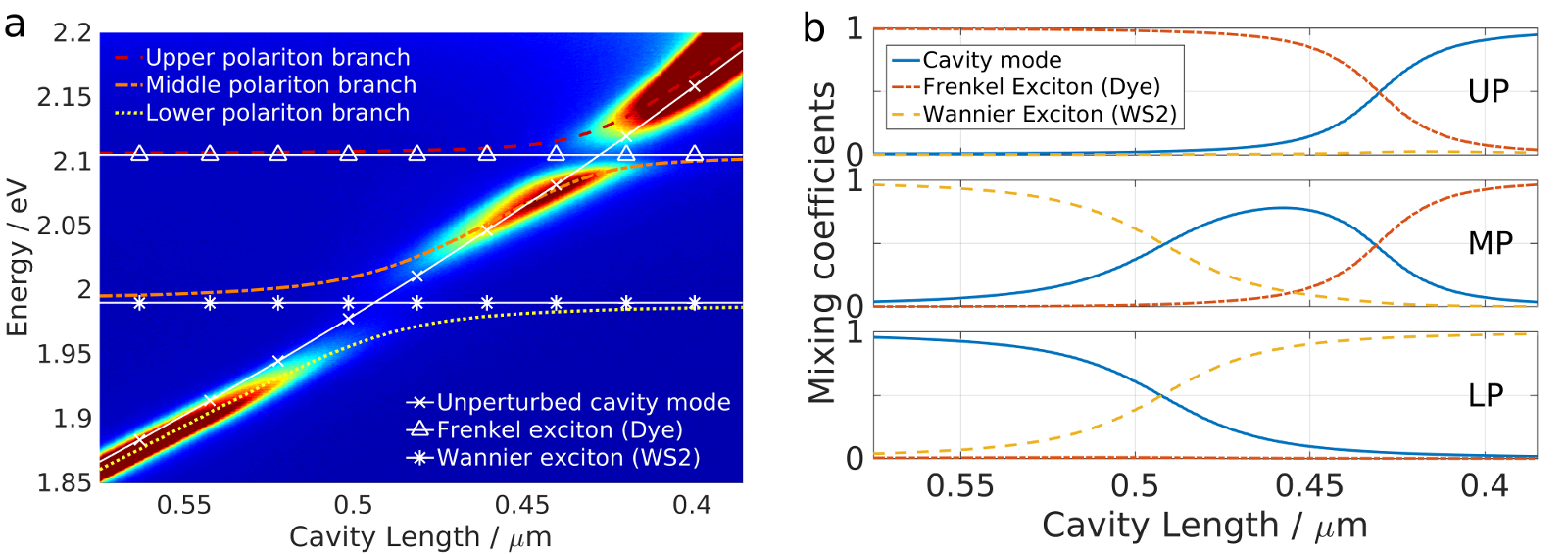}
\caption[flushleft]{\textbf{Hybrid polariton formation with lower concentration dye.} a) Successive transmission spectra as the cavity length is decreased from left to right, symbols and lines as in main text. b) Hopfield coefficients as presented in main text.
 \label{figS1}} 
\end{figure}

\begin{figure}[h!]
\centering
\includegraphics[width=0.95\textwidth]{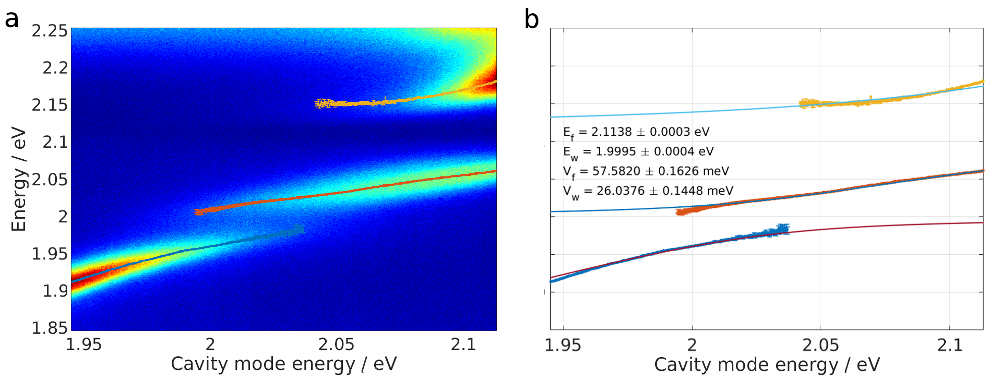}
\caption[flushleft]{\textbf{Peak and dispersion fitting} a) Fitted transmission peaks with error bars superimposed on raw transmission data for different cavity mode energies. Each peak is fitted with a Lorentzian profile. b) Dispersion as given by Eq. 2 fitted to the peak positions presented in a. The fit is performed after obtaining three separate equations for UP, MP and LP from diagonalisation of Eq. 2 and proceeding with a non-linear least squares algorithm with shared parameters $E_{\rm{f}}$, $E_{\rm{w}}$, $V_{\rm{f}}$ and $V_{\rm{w}}$. \label{figS8} }
\end{figure}

\begin{figure}[h!]
\centering
\includegraphics[width=0.8\textwidth]{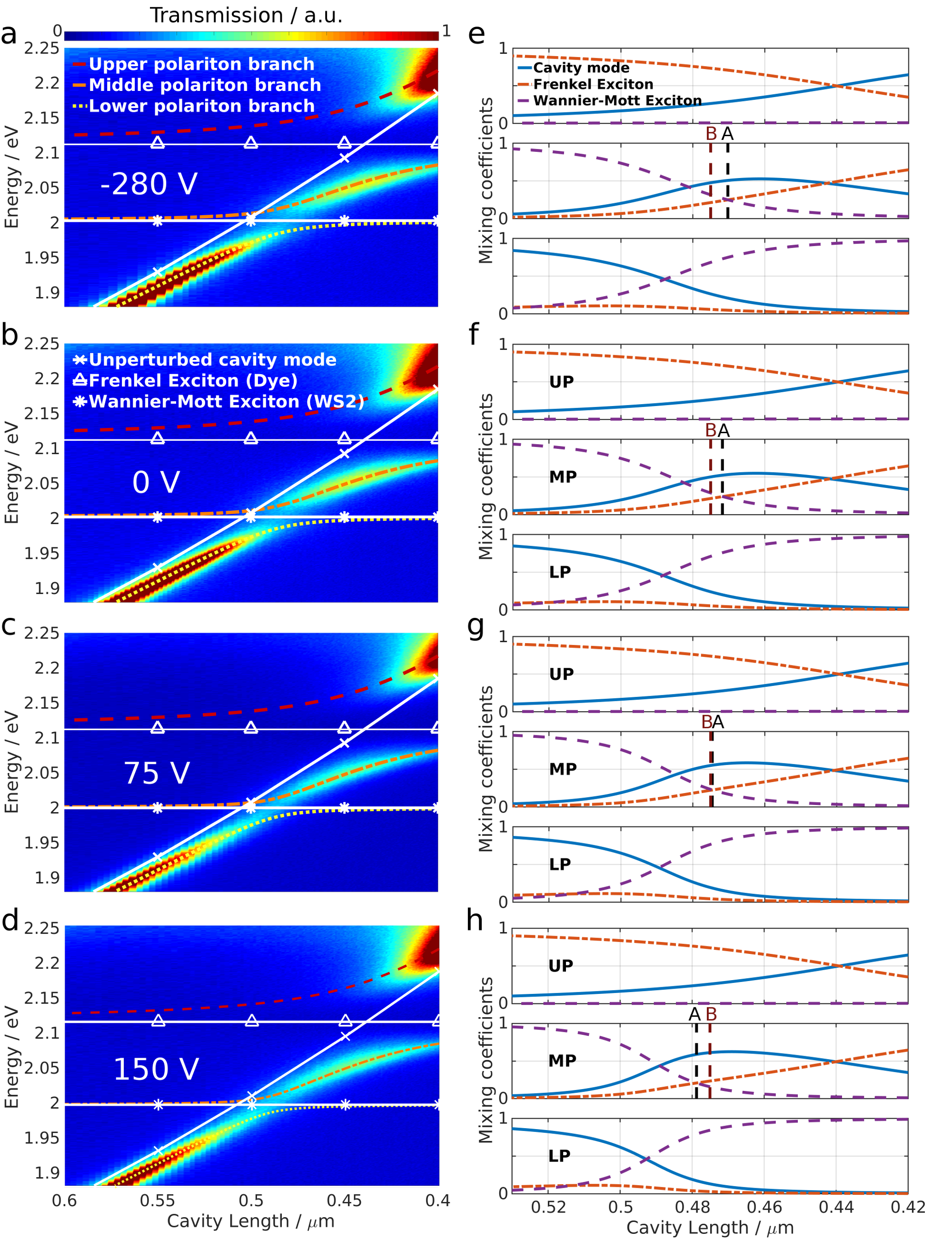}
\caption[flushleft]{\textbf{Electrical tuning of hybrid polaritons on a second WS$_2$ flake.} a-d) Successive transmission spectra of hybrid WS$_2$-TDBC microcavity for decreasing cavity length from left to right and different applied voltages of \SI{-280}{\volt}, \SI{0}{\volt}, \SI{75}{\volt} and \SI{150}{\volt} for a, b, c and d respectively. The white, continuous lines correspond to the uncoupled energies of Frenkel-exciton (TDBC, triangles), Wannier-Mott exciton (WS$_2$, stars) and cavity mode (crosses). The dashed lines in colour show the dispersion for the coupled system consisting of the three polariton branches, lower polariton (LP, yellow, - -), middle polariton (MP, orange, $-$ -) and upper polariton (UP, red, $-$ $-$). e-h) Photonic (cavity mode, blue, continuous), Frenkel-excitonic (TDBC, red,$-$ -) and Wannier-Mott-excitonic (WS$_2$, purple, - -) contribution to the three polariton branches LP, MP and UP for the dispersions plotted respectively above. Two points A (black dashed line) and B (red dashed line) mark cavity lengths at which: A) Frenkel- and Wannier-Mott-exciton contribution to the middle polariton branch is equal, B) the cavity length is $L=$~\SI{0.475}{\micro\meter}. \label{figS3} } 
\end{figure}

\begin{figure}[h!]
\centering
\includegraphics[width=0.95\textwidth]{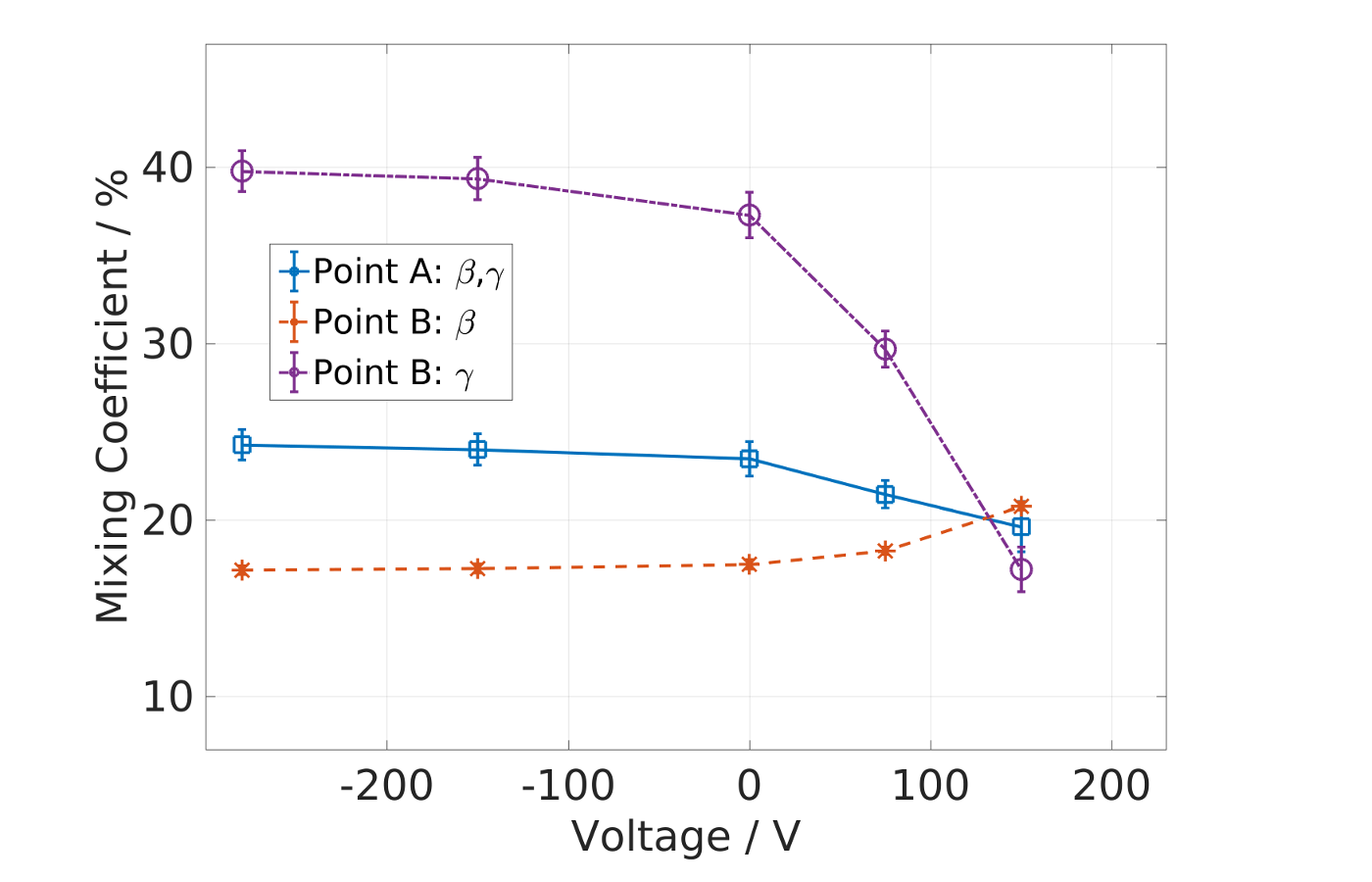}
\caption[flushleft]{\textbf{Electrical control over composition of polaritons for a second WS$_2$ flake.} Frenkel- and Wannier-Mott-exciton fraction $\beta$ and $\gamma$ in hybrid polariton state for different applied voltages (colour, symbols, left ordinate) and absorbance peak position of WS$_2$ outside the cavity (black, dashed, right ordinate). The exciton fractions are shown for two points, where point A corresponds to the cavity length at which $\beta = \gamma$ (maximal mixing) and point B to a fixed cavity length of $L=$~\SI{0.475}{\micro\meter}. \label{figS4} }
\end{figure}
\clearpage

\section{D\MakeLowercase{etails of experimental setup}}
\subsection{Sample preparation}
In the following section we describe the two mirrors forming the cavity and the absorbing material within. more details on the measurements are given in the next section. The first mirror consists of 10 pair DBR of SiO$_2$, TiO$_2$ with refractive indices 1.45 and 2.05 respectively deposited on a \SI{0.5}{\milli\meter} thick, flat Silica substrate. The stopband of this mirror is centered around $\lambda = $ \SI{637}{\nano\meter} and the reflectivity at that wavelength is 99.7$\%$. The second, smaller mirror is produced by removing large areas of a flat silica substrate with a dicer to create a 200 $\times$ 300 \SI{}{\micro\meter}$^2$ plinth made semi-reflective by thermally evaporating a \SI{50}{\nano\meter} thick silver layer. 
To enable electrical control within the cavity, silver electrodes with a width and a spacing of \SI{90}{\micro\meter} are thermally evaporated on top of the DBR. This process is facilitated by masking a region of the DBR with a laser processed foil defining the electrode shape. The thickness of these electrodes is similar to the thickness of the silver layer on the small opposing mirror, approximately \SI{50}{\nano\meter}. The WS$_2$ flakes are grown as described in \cite{rong_controlling_2014} and transferred onto the dielectric mirror stack, which has a low refractive-index terminated layer to provide an anti-node of the electric field at the mirror surface and thus optimal coupling to the monolayer. This transfer is facilitated by spin-coating a thick layer of PMMA onto the as-grown WS$_2$ flakes on their native SiO$_2$ substrate. After etching the substrate away the PMMA stays connected with the WS$_2$ flakes and can be fished out and cut manually. The resulting PMMA section is then manually positioned over the DBR region holding the electrodes and baked at 100$^\circ$C for one hour. After this the PMMA is removed by putting the sample into an acetone bath for about \SI{15}{\minute}. The distribution of WS$_2$ flakes relative to the electrodes is random, the average size of a WS$_2$ flake being \SI{80}{\micro\meter}, with 90$\%$ of the flakes overlapping partially with the silver electrodes (see Fig.~\ref{figS7}). The results presented above were obtained from a flake which was not in contact with the electrodes, thus ensuring purely electrostatic tuning. The electrodes have a length of $\approx$~\SI{1}{\centi\meter} and were connected to thin wires with conducting silver paint at the respective ends. A cavity mode in the region of these electrodes would be spectrally broader and the number of round-trips would be reduced due to the additional absorption of the second silver interface. For the data that we present in the manuscript the effect that such broadening would have can be estimated: \newline Given the cavity length of around $500$ nm and the reflectivity of the silver mirror $R = 0.95$, we can equate an effective mode area $A = \frac{\pi L \lambda}{1-R} \approx 19.5$ $\mu$m$^2$, as derived in \cite{ujihara_spontaneous_1991}.  This area translates to a mode radius of $r \approx 2.5$ $\mu$m. Since the region of the sample from which we obtained the results is more than $15$ $\mu$m from the next silver electrode, the effect from the electrode on the cavity mode is negligible. \newline

\noindent
The organic dye was introduced on the opposing mirror by dissolving J-aggregated 1,1\unichar{8242}-diethyl-3,3\unichar{8242}-di(4-sulfobutyl)-5,5\unichar{8242},6,6\unichar{8242}-tetrachlorobenzimidazolocarbocyanine (TDBC) in an aqueous solution with 5 weight percent gelatine. The solution was then spin coated onto the small silver mirror, giving a polymer, dye layer of approximately \SI{300}{\nano\meter}. 
The j-aggregate exciton energy can be shifted in an electrical field through the Stark effect, but it requires field strengths of about $10^6$ V/cm for a 20 nm shift (as determined through electroabsorption measurements - unpublished), whereas the field strengths that we obtained in our experiment were about $2.33 \times 10^4$ V/cm. Our data gives experimental evidence of this, since the Frenkel exciton energy remained unchanged (within the uncertainty of the Lorentzian lineshape fit) for any applied electric field.

\subsection{Optical measurements}
The small silver mirror is mounted on a three-dimensional piezo actuated stage, which makes electric positioning relative to the WS$_2$ flake possible. To initialise the cavity, white light from a light emitting diode is shone through the mirrors while reducing the separation. With the help of Fabry-Perot fringes visible for small mirror separations ($L < $ \SI{20}{\micro\meter}) both surfaces are made parallel within \SI{150}{\micro\radian}.
Optical access to the sample is given by a standard $\times 10$ objective lens and the collected light is focused on an Andor combined spectrograph/CCD for analysis.
The electrodes were pairwise connected to a Keithley 2400 to apply voltages and monitor the current. Care was taken that for the datasets presented, the full voltage was applied and the respective current limits were not exceeded.

\section{C\MakeLowercase{omponents of hybrid polariton system}}
In the following we present a characterisation of the individual components of the hybrid system. Fig.~\ref{figS2}a shows transmission spectra of a microcavity with only a WS$_2$ flake  for different cavity lengths. As the cavity length is decreased from left to right, the cavity mode energy increases and traverses the exciton energy. As both components strongly couple to each other, an avoided level crossing with characteristic Rabi splitting is visible. The coloured, dashed lines show the dispersion of the lower and upper polariton branch. Fig.~\ref{figS2}b shows the photonic and excitonic fractions of the resulting polariton states, which are known as the Hopfield coefficients, as introduced in the main text. We have published a study on the properties of this open cavity polariton system \cite{flatten_room-temperature_2016}. 

A strongly coupled system on the basis of organic dyes is well known in the literature \cite{coles_imaging_2013,coles_polariton-mediated_2014}. Fig.~\ref{figS6} shows the Rabi splitting for a fixed cavity length as a function of the concentration of the dye TDBC. When plotted as a function of the square root of the concentration the linear relationship is visible. By varying the concentration of TDBC in the aquaeous solution it is possible to optimise the hybridisation between WS$_2$ and organic excitons.  

Fig.~\ref{figS1}a shows transmission spectra of a hybrid microcavity system with a lower concentrated TDBC layer compared to the data presented before in the manuscript. Symbols and lines are defined as previously in the main text. As the strength of the interaction between dye and cavity mode is reduced, the photonic fraction in the middle polariton branch is increased and the hybridisation between both excitons only occurs marginally ($\beta^2 = \gamma^2 \approx 10 \%$ for a cavity length of $L =$~\SI{0.455}{\micro\meter}). By variation of the concentration of the dye it is thus possible to control the composition of the polariton branches, in particular the middle polariton branch. When relying on the photonic part of a polariton state for detection, a reduced overlap between both excitons is desirably - hence the choice of concentration in the main text. For applications, where the exciton hybridisation is the only important parameter, a larger dye concentration might be desirable.

\section{E\MakeLowercase{lectrical control of \MakeUppercase{WS}$_2$ absorption profile}}
The occurence of trion (or charged exciton) states in atomically flat TMDCs has been reported previously \cite{chernikov_electrical_2015,singh_trion_2016}. By applying an electric field to deplete a region of the material from electrons, it is possible to change the Fermi level and thus the spectral weight and position of both the neutral exciton (X$^0$) and charged 
exciton (X$^-$) states
\cite{ross_electrical_2013,zhu_exciton_2015,chernikov_electrical_2015,sidler_fermi_2016}. Such electrical control can thus be used to vary the absorption profile. Fig.~\ref{figS5} shows the absorbance of the WS$_2$ for different applied voltages. As the Fermi level is raised for increasing positive biases the spectral weight of the absorption profile shifts towards the charged exciton X$^-$.

As Fig. 2(a) in Ref. \cite{zhu_exciton_2015} shows, the trion can be quenched by applying a negative gate voltage. The zero bias level of weight in neutral and charged exciton is dependent on the intrinsic doping of the WS$_2$ flake, which is thought to be a result of the growth and transfer method. On our CVD grown samples we see variations of X$^-$ contribution on one flake and even larger variations when comparing multiple flakes. In general the X/X$^-$ ratio found for our sample is similar to bias values between -30 V and 0 V in Fig. 2(a) in Ref. \cite{zhu_exciton_2015}. To demonstrate the strong coupling to a cavity mode we chose a region with minimal trion contribution (as visible in the absorption lineshape in Fig. 1 d) for negative applied voltages). The reasoning behind this was, that with more trionic contribution the polariton linewidth increases, the asymmetry between upper and lower polariton branch increases and the splitting is less obvious.

\section{F\MakeLowercase{itting procedure and error propagation}}
Each transmission data set is aquired and analysed by the following procedure: The sequence of transmission spectra is aquired by sweeping the voltage of the piezo microactuator, varying the cavity length. The actual cavity length for each frame is obtained by fitting a Lorentzian profile to the unperturbed cavity mode for either the same longitudinal mode (with index $q=4$ and for energies below $E = $\SI{1.85}{\milli\electronvolt} or the the next one (with index $q+1 = 5$, if the $q=4$ mode has energies above $E = $\SI{1.85}{\milli\electronvolt}). The value of $q$ can be obtained for each dataset from the free spectral range, which then allows the absolute cavity length to be expressed as $L_{cav} = \frac{qhc}{2E}$, where $E$ is the mode energy. Each frame is fitted with Lorentzian lineshape peaks to obtain the position of the individual polariton branches (Fig.~\ref{figS8}a). The analytic form of the three equations for UP, MP and LP is found by diagonalising Eq. 2. The three expressions are simultanuously fitted to the obtained peaks with a nonlinear least squares algorithm (Fig.~\ref{figS8}b). In a first round, the four parameters $(E_{\rm{f}}$, $E_{\rm{w}}$, $V_{\rm{f}}, V_{\rm{w}})$ are shared in the fitting procedure and obtained as parameters after the fit. In a second round the parameters governing the position and the interaction strength of the dye $(E_{\rm{f}}$, $V_{\rm{f}})$ are fixed to a common value obtained by averaging their respective values from the first round. The covariance matrix $M_{\rm{cov}}$ is obtained for each fit. Through diagonalisation of Eq. 2 we obtain the eigenvectors, whose components represent the mixing coefficients after proper normalisation. The respective values for point A and B of these values is directly obtained by plugging in the parameters found above. The uncertainty for each of the coefficients is obtained by constructing the Jacobian $J^f_{ij}$ = $\frac{\partial f_i}{\partial x_j}$, where $f_i$ is the respective expression for the polariton fraction and $x_j$ is one of the parameters found above. The uncertainties $u$ can now be calculated from $u = \sqrt{\rm{diag}(J M_{cov} J^t)}$.

\section{E\MakeLowercase{lectrical control on different \MakeUppercase{WS}$_2$ flake}}
Complementary to the data set shown in the main text, we include another dataset showing the electrical control of the hybridisation. Fig.~\ref{figS3}a-d show transmission spectra at a different point on the sample for different applied voltages. The symbols and lines are defined as in the main text and the figure caption. Fig.~\ref{figS3}e-h show the photonic (blue, continuous), Wannier-Mott (purple, dashed) and Frenkel-excitonic (red,dashed) fractions of the three polariton branches corresponding to the dispersion shown to the left in a-d.

Fig.~\ref{figS4} presents the summary of the electrically controlled hybridisation analogous to Fig.~3 in the main text for the second dataset.

\clearpage

\putbib

\end{bibunit}


\begin{thebibliography}{10}

\bibitem{weisbuch_observation_of_the}
C.~Weisbuch, M.~Nishioka, A.~Ishikawa, and Y.~Arakawa, ``Observation of the
  coupled exciton-photon mode splitting in a semiconductor quantum
  microcavity,'' {\em Phys. Rev. Lett.}, vol.~69, pp.~3314--3317, Dec 1992.

\bibitem{lidzey_strong_exciton_photon}
D.~G. Lidzey, D.~D.~C. Bradley, M.~S. Skolnick, T.~Virgili, S.~Walker, and
  D.~M. Whittaker, ``Strong exciton-photon coupling in an organic semiconductor
  microcavity,'' {\em Nature}, vol.~395, pp.~53--55, 1998.

\bibitem{dang_stimulation_of_polariton}
L.~S. Dang, D.~Heger, R.~Andr\'e, F.~B\oe{}uf, and R.~Romestain, ``Stimulation
  of polariton photoluminescence in semiconductor microcavity,'' {\em Phys.
  Rev. Lett.}, vol.~81, pp.~3920--3923, Nov 1998.

\bibitem{senellart_nonlinear_emission_of}
P.~Senellart and J.~Bloch, ``Nonlinear emission of microcavity polaritons in
  the low density regime,'' {\em Phys. Rev. Lett.}, vol.~82, pp.~1233--1236,
  Feb 1999.

\bibitem{kasprzak_bose_einstein_condensation}
J.~Kasprzak, M.~Richard, S.~Kundermann, A.~Baas, P.~Jeambrun, J.~M.~J. Keeling,
  F.~M. Marchetti, M.~H. Szymanska, R.~Andre, J.~L. Staehli, V.~Savona, P.~B.
  Littlewood, B.~Deveaud, and L.~S. Dang, ``Bose-einstein condensation of
  exciton polaritons,'' {\em Nature}, vol.~443, p.~409, 2006.

\bibitem{plumhof_room-temperature_2014}
J.~D. Plumhof, T.~St{\"{o}}ferle, L.~Mai, U.~Scherf, and R.~F. Mahrt,
  ``Room-temperature {Bose}-{Einstein} condensation of cavity
  exciton–polaritons in a polymer,'' {\em Nature Materials}, vol.~13,
  pp.~247--252, Mar. 2014.

\bibitem{amo_superfluidity_of_polaritons}
A.~Amo, J.~Lefrere, S.~Pigeon, C.~Adrados, C.~Ciuti, I.~Carusotto, R.~Houdre,
  E.~Giacobino, and A.~Bramati, ``Superfluidity of polaritons in semiconductor
  microcavities,'' {\em Nature Physics}, vol.~5, p.~805, 2009.

\bibitem{schneider_electrically_2013}
C.~Schneider, A.~Rahimi-Iman, N.~Y. Kim, J.~Fischer, I.~G. Savenko, M.~Amthor,
  M.~Lermer, A.~Wolf, L.~Worschech, V.~D. Kulakovskii, I.~A. Shelykh, M.~Kamp,
  S.~Reitzenstein, A.~Forchel, Y.~Yamamoto, and S.~H{\"{o}}fling, ``An
  electrically pumped polariton laser,'' {\em Nature}, vol.~497, pp.~348--352,
  May 2013.

\bibitem{rodriguez_interaction-induced_2016}
S.~R.~K. Rodriguez, A.~Amo, I.~Sagnes, L.~L. Gratiet, E.~Galopin,
  A.~Lema{\^{i}}tre, and J.~Bloch, ``Interaction-induced hopping phase in
  driven-dissipative coupled photonic microcavities,'' {\em Nature
  Communications}, vol.~7, p.~11887, June 2016.

\bibitem{lidzey_room_temperature_polariton}
D.~G. Lidzey, D.~D.~C. Bradley, T.~Virgili, A.~Armitage, M.~S. Skolnick, and
  S.~Walker, ``Room temperature polariton emission from strongly coupled
  organic semiconductor microcavities,'' {\em Phys. Rev. Lett.}, vol.~82,
  pp.~3316--3319, Apr 1999.

\bibitem{zhu_exciton_2015}
B.~Zhu, X.~Chen, and X.~Cui, ``Exciton {Binding} {Energy} of {Monolayer}
  {WS}2,'' {\em Scientific Reports}, vol.~5, p.~9218, Mar. 2015.

\bibitem{schmidt_electronic_2015}
H.~Schmidt, F.~Giustiniano, and G.~Eda, ``Electronic transport properties of
  transition metal dichalcogenide field-effect devices: surface and interface
  effects,'' {\em Chemical Society Reviews}, vol.~44, pp.~7715--7736, Oct.
  2015.

\bibitem{liu_electronic_2015}
G.-B. Liu, D.~Xiao, Y.~Yao, X.~Xu, and W.~Yao, ``Electronic structures and
  theoretical modelling of two-dimensional group-{VIB} transition metal
  dichalcogenides,'' {\em Chem. Soc. Rev.}, vol.~44, pp.~2643--2663, Apr. 2015.

\bibitem{ross_electrical_2013}
J.~S. Ross, S.~Wu, H.~Yu, N.~J. Ghimire, A.~M. Jones, G.~Aivazian, J.~Yan,
  D.~G. Mandrus, D.~Xiao, W.~Yao, and X.~Xu, ``Electrical control of neutral
  and charged excitons in a monolayer semiconductor,'' {\em Nature
  Communications}, vol.~4, p.~1474, Feb. 2013.

\bibitem{allain_electrical_2015}
A.~Allain, J.~Kang, K.~Banerjee, and A.~Kis, ``Electrical contacts to
  two-dimensional semiconductors,'' {\em Nature Materials}, vol.~14,
  pp.~1195--1205, Dec. 2015.

\bibitem{klots_probing_2014}
A.~R. Klots, A.~K.~M. Newaz, B.~Wang, D.~Prasai, H.~Krzyzanowska, J.~Lin,
  D.~Caudel, N.~J. Ghimire, J.~Yan, B.~L. Ivanov, K.~A. Velizhanin, A.~Burger,
  D.~G. Mandrus, N.~H. Tolk, S.~T. Pantelides, and K.~I. Bolotin, ``Probing
  excitonic states in suspended two-dimensional semiconductors by photocurrent
  spectroscopy,'' {\em Scientific Reports}, vol.~4, p.~6608, Oct. 2014.

\bibitem{holmes_strong_coupling_and}
R.~J. Holmes, S.~K{\'{e}}na-Cohen, V.~M. Menon, and S.~R. Forrest, ``Strong
  coupling and hybridization of frenkel and wannier-mott excitons in an
  organic-inorganic optical microcavity,'' {\em Phys. Rev. B}, vol.~74,
  p.~235211, Dec 2006.

\bibitem{wenus_hybrid_organic_inorganic}
J.~Wenus, R.~Parashkov, S.~Ceccarelli, A.~Brehier, J.-S. Lauret, M.~S.
  Skolnick, E.~Deleporte, and D.~G. Lidzey, ``Hybrid organic-inorganic
  exciton-polaritons in a strongly coupled microcavity,'' {\em Phys. Rev. B},
  vol.~74, p.~235212, Dec 2006.

\bibitem{rong_controlling_2014}
Y.~Rong, Y.~Fan, A.~Leen~Koh, A.~W. Robertson, K.~He, S.~Wang, H.~Tan,
  R.~Sinclair, and J.~H. Warner, ``Controlling sulphur precursor addition for
  large single crystal domains of {WS}$_2$,'' {\em Nanoscale}, vol.~6,
  pp.~12096--12103, Sept. 2014.

\bibitem{flatten_room-temperature_2016}
L.~C. Flatten, Z.~He, D.~M. Coles, A.~A.~P. Trichet, A.~W. Powell, R.~A.
  Taylor, J.~H. Warner, and J.~M. Smith, ``Room-temperature exciton-polaritons
  with two-dimensional {WS}2,'' {\em Scientific Reports}, vol.~6, p.~33134,
  Sept. 2016.

\bibitem{mak_tightly_2013}
K.~F. Mak, K.~He, C.~Lee, G.~H. Lee, J.~Hone, T.~F. Heinz, and J.~Shan,
  ``Tightly bound trions in monolayer {MoS}2,'' {\em Nature Materials},
  vol.~12, pp.~207--211, Mar. 2013.

\bibitem{plechinger_identification_2015}
G.~Plechinger, P.~Nagler, J.~Kraus, N.~Paradiso, C.~Strunk, C.~Sch{\"{u}}ller,
  and T.~Korn, ``Identification of excitons, trions and biexcitons in
  single-layer {WS}2,'' {\em Phys. Status Solidi RRL}, vol.~9, pp.~457--461,
  Aug. 2015.

\bibitem{chernikov_electrical_2015}
A.~Chernikov, A.~M. van~der Zande, H.~M. Hill, A.~F. Rigosi, A.~Velauthapillai,
  J.~Hone, and T.~F. Heinz, ``Electrical {Tuning} of {Exciton} {Binding}
  {Energies} in {Monolayer} {WS}$_2$,'' {\em Phys. Rev. Lett.}, vol.~115,
  p.~126802, Sept. 2015.

\bibitem{sidler_fermi_2016}
M.~Sidler, P.~Back, O.~Cotlet, A.~Srivastava, T.~Fink, M.~Kroner, E.~Demler,
  and A.~Imamoglu, ``Fermi polaron-polaritons in charge-tunable atomically thin
  semiconductors,'' {\em Nature Physics}, Oct. 2016.

\bibitem{baeten_many-body_2015}
M.~Baeten and M.~Wouters, ``Many-body effects of a two-dimensional electron gas
  on trion-polaritons,'' {\em Physical Review B}, vol.~91, p.~115313, Mar.
  2015.

\bibitem{portella-oberli_nonlinear_2003}
M.~T. Portella-Oberli, V.~Ciulin, M.~Kutrowski, T.~Wojtowicz, and B.~Deveaud,
  ``Nonlinear optical dynamics of excitons and trions,'' {\em Phys. Status
  Solidi B}, vol.~238, pp.~513--516, Aug. 2003.

\bibitem{fraser_physics_2016}
M.~D. Fraser, S.~H{\"{o}}fling, and Y.~Yamamoto, ``Physics and applications of
  exciton-polariton lasers,'' {\em Nat Mater}, vol.~15, pp.~1049--1052, Oct.
  2016.

\end{thebibliography}


\begin{thebibliography}{10}

\bibitem{rong_controlling_2014}
Y.~Rong, Y.~Fan, A.~Leen~Koh, A.~W. Robertson, K.~He, S.~Wang, H.~Tan,
  R.~Sinclair, and J.~H. Warner, ``Controlling sulphur precursor addition for
  large single crystal domains of {WS}$_2$,'' {\em Nanoscale}, vol.~6,
  pp.~12096--12103, Sept. 2014.

\bibitem{ujihara_spontaneous_1991}
K.~Ujihara, ``Spontaneous {Emission} and the {Concept} of {Effective} {Area} in
  a {Very} {Short} {Optical} {Cavity} with {Plane}-{Parallel} {Dielectric}
  {Mirrors},'' {\em Japanese Journal of Applied Physics}, vol.~30,
  pp.~L901--L903, May 1991.

\bibitem{flatten_room-temperature_2016}
L.~C. Flatten, Z.~He, D.~M. Coles, A.~A.~P. Trichet, A.~W. Powell, R.~A.
  Taylor, J.~H. Warner, and J.~M. Smith, ``Room-temperature exciton-polaritons
  with two-dimensional {WS}2,'' {\em Scientific Reports}, vol.~6, p.~33134,
  Sept. 2016.

\bibitem{coles_imaging_2013}
D.~M. Coles, R.~T. Grant, D.~G. Lidzey, C.~Clark, and P.~G. Lagoudakis,
  ``Imaging the polariton relaxation bottleneck in strongly coupled organic
  semiconductor microcavities,'' {\em Physical Review B}, vol.~88, p.~121303,
  Sept. 2013.

\bibitem{coles_polariton-mediated_2014}
D.~M. Coles, N.~Somaschi, P.~Michetti, C.~Clark, P.~G. Lagoudakis, P.~G.
  Savvidis, and D.~G. Lidzey, ``Polariton-mediated energy transfer between
  organic dyes in a strongly coupled optical microcavity,'' {\em Nature
  Materials}, vol.~13, pp.~712--719, July 2014.

\bibitem{chernikov_electrical_2015}
A.~Chernikov, A.~M. van~der Zande, H.~M. Hill, A.~F. Rigosi, A.~Velauthapillai,
  J.~Hone, and T.~F. Heinz, ``Electrical {Tuning} of {Exciton} {Binding}
  {Energies} in {Monolayer} {WS}$_2$,'' {\em Phys. Rev. Lett.}, vol.~115,
  p.~126802, Sept. 2015.

\bibitem{singh_trion_2016}
A.~Singh, G.~Moody, K.~Tran, M.~E. Scott, V.~Overbeck, G.~Bergh{\"{a}}user,
  J.~Schaibley, E.~J. Seifert, D.~Pleskot, N.~M. Gabor, J.~Yan, D.~G. Mandrus,
  M.~Richter, E.~Malic, X.~Xu, and X.~Li, ``Trion formation dynamics in
  monolayer transition metal dichalcogenides,'' {\em Physical Review B},
  vol.~93, p.~041401, Jan. 2016.

\bibitem{ross_electrical_2013}
J.~S. Ross, S.~Wu, H.~Yu, N.~J. Ghimire, A.~M. Jones, G.~Aivazian, J.~Yan,
  D.~G. Mandrus, D.~Xiao, W.~Yao, and X.~Xu, ``Electrical control of neutral
  and charged excitons in a monolayer semiconductor,'' {\em Nature
  Communications}, vol.~4, p.~1474, Feb. 2013.

\bibitem{zhu_exciton_2015}
B.~Zhu, X.~Chen, and X.~Cui, ``Exciton {Binding} {Energy} of {Monolayer}
  {WS}2,'' {\em Scientific Reports}, vol.~5, p.~9218, Mar. 2015.

\bibitem{sidler_fermi_2016}
M.~Sidler, P.~Back, O.~Cotlet, A.~Srivastava, T.~Fink, M.~Kroner, E.~Demler,
  and A.~Imamoglu, ``Fermi polaron-polaritons in charge-tunable atomically thin
  semiconductors,'' {\em Nature Physics}, Oct. 2016.

\end{thebibliography}
\end{document}